# The Enhancement of Communication Technologies and Networks for Smart Grid Applications

**Saida Elyengui[1], Riadh Bouhouchi[2], Tahar Ezzedine[3]**

[1,2,3] Université de Tunis El Manar, Ecole Nationale d'Ingénieurs de Tunis, LR99ES21 Laboratoire de Systèmes de Communications, 1002, Tunis, Tunisie.

**Abstract:** *The current electrical grid is perhaps the greatest engineering achievement of the 20th century. However, it is increasingly outdated and overburdened, leading to costly blackouts and burnouts. For this and various other reasons, transformation efforts are underway to make the current electrical grid smarter. A reliable, universal and secure communication infrastructure is mandatory for the implementation and deployment of the future smart grid. A special interest is given to the design of efficient and robust network architecture capable of managing operation and control of the next generation power grid. For this purpose new wired and wireless technologies are emerging in addition to the formerly applied to help upgrade the current power grid. In this paper we will give an overview of smart grid reference model, and a comprehensive survey of the available networks for the smart grid and a critical review of the progress of wired and wireless communication technologies for smart grid communication infrastructure. And we propose end to end communication architecture for Home Area Networks (HANs), Neighborhood Area Networks (NANs) and Wide Area Networks (WANs) for smart grid applications. We believe that this work will provide appreciated insights for the novices who would like to follow related research in the SG domain.*

## 1. Introduction

The researches in smart grid domain are enhanced by dramatic economic losses and repetitive electrical systems blackouts and failures. Advanced communication and networking technologies will be incorporated in the future electrical power system in order to make the grid more reliable, secure and sustainable. In smart grid communication the network is essential to connect intelligent electronic devices in distributed locations in order to establish stable bidirectional exchange of data flow, control and monitoring instructions between them and utilities control centers. In addition a strong, reliable, secure and robust communication infrastructure is required to gather, collect, and combine data provided by smart meters, computers, sensors and electrical vehicles or any electrical smart devices connected to the grid to help providing better power quality and efficient delivery.

This paper is organized as follows, In Section II we pointed some related work that have been done in communication and networking for smart grid, a brief overview of smart grid conceptual architecture and reference model is presented in section III. In Section VI we propose whole study of present communication networks for smart grid applications. In section V and VI we perform a review of most important wired and wireless communication technologies for the development of smart grid and we perform a comparative study by describing advantages and disadvantages and proposing a bench of SG applications for each technology .In Section VII we advocate an end to end communication architecture, in section VIII we list some of communication challenges for smart grid and finally in section IX we conclude and give some of our future directions.

## 2. Related Work

In this section, we highlight some of the work done in the communication and networking for smart grid. In [1] the authors presented a smart grid communication networking architecture and communication technologies, they thoroughly discussed power line communication and wireless technologies. In another research work, [2] the authors identified communication architecture and functional requirement and introduced HANs, FANs and WANs networks. In addition the authors focused on communication functionalities and requirements. Moreover, in [3] the authors depicted the conceptual architecture for smart power grid; they presented also different communication technologies for smart grids. To the best of our knowledge, this work will provide a new comprehensive study of the most important evolution of communication and networking technologies for smart grids and deliver detailed end to end communication network architecture.

## 3. Smart Grid Reference Model

We are actually facing several challenges in order to build smart grid communication infrastructure such as scalability, interoperability, security, customer's privacy, and smart meter infrastructure incorporation with different utility companies and consumers facilities [1]. According to Smart Grid Interoperability Standards Roadmap [4] proposed by NIST the American National

Institute of Standards and Technology, the conceptual architecture for smart grid is composed of seven big domains as illustrated by figure 1 , these domains describes  the scope of smart grid infrastructure .

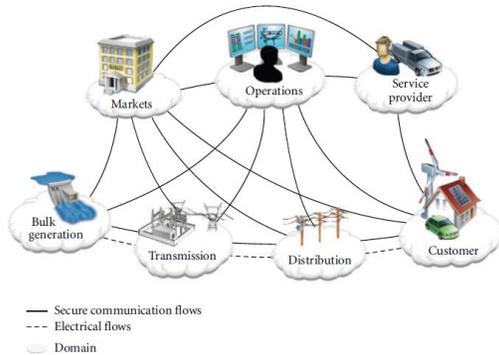

Source : National Institute of Standards and Technology framework and roadmap for smart grid interoperability standards.

**Figure 1** NIST framework for smart grid

All these functional domains have different inter and intra domain communications, the "consumer" domain is the user of electricity domain such as domestic, industrial, commercial or utilities. In smart grid approach the consumers can produce, store and use the electrical power. The "market" domain refers to power market operators. The "operation" domain deal with power supply management. "Service provider" points service utilities companies providing customers with electrical power. ''Bulk Generation'', ''Transmission'' and ''Distribution'' refers to generation, storage, transmission and distribution of power to customers. One of the key elements of smart grid successful operation is the interconnection of these seven domains. For this purpose new communication technologies are required to integrate these domains with interoperability standards, some of these priority technologies will be considered in following section in this work.

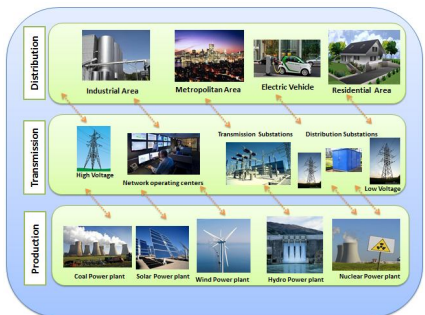

**Figure 2** Smart Grid tired high level Architecture

The authors in [7] describes all the power production, generation, transmission and distribution to customers using different production mechanisms and various generation plants types as fixed or mobiles depending on precise and particular as illustrated by the figure 2. The production domain is composed of a mixture of nuclear, solar, coal, wind  or hydro power plant .The transmission domains  is  managed  by  huge  number  of  network operating centers and substations, a large number of power lines deliver the electricity to distribution domain. Finally in the distribution domain a sum of complex networks topologies delivers electrical power to residential areas, rural farms, metropolitan areas, and industrial areas for consumption. In our study we focus in the distribution domain of the smart grid. Especially, our interest is given to the communication networks in the distribution domains of the smart grid; the home area networks (HAN), the neighbor area networks (NAN) and the wide area networks (WAN). These networks are very important for data flows transportation between end consumers and utilities.

## 4. Smart grid communication networks

There has been several survey and research papers done for SG communication lately [2, 12, 13, 14, and 15], that evaluate the works and the features of communication and networking infrastructure for smart grid systems and applications.

The communication networks for smart grid systems use a big range of communication technologies from wired, wireless and hybrid networks technologies. The actual electrical grid has already a communications networks supporting its operations between substations and control centers but this network is expensive, uncompromising and insufficient because it covers only generation and transmission segments. In the prospect smart grid approach we are aiming to cover all networks segments and especially the distribution segment. Therefore a whole new adequate and extended communication network is needed to support  and back up new SG applications and systems and to meet the upcoming demands [12]. We believe that communication infrastructure for smart grid will be a hybrid mesh composed of various networking topologies and technologies [16] as illustrated in figure 3.

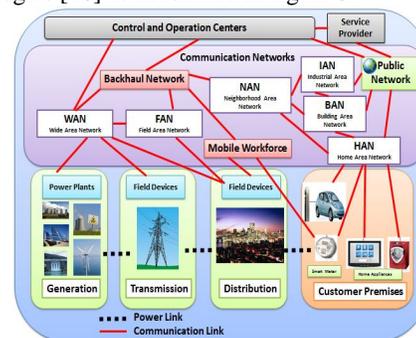

**Figure 3** Multi-tier communication networks for Smart Grid

However in order to monitor, control and have bi-directional data  flows between end devices and smart grid utilities a highly reliable , integrated communication network should cover all the SG domains. In the following section we will focus in the three most important SG communication networks to permit a better understanding of smart grid communication networks.

*Home Area Network (HAN)*
Located in customer domain, Home area network offers access to in-home smart devices and appliances. IED send data readings over HAN to AMI applications threw the home smart meter or threw the residential gateway. The HAN give also to home automation networks different services like home motoring and control, demand response applications allowing efficient power management and user comfort [14]. We can find various sensors categories in home automation networks, such as light control sensors, temperature and humidity sensors, remote care and control sensors, motions sensors, security and safety sensors etc.
BAN and IAN networks refer to HAN parallel networks when implemented respectively in business/buildings or industrial areas. The most suitable communication technology in HAN/BAN/IAN networks is wireless technology since its ease of implementation for a big number of nodes, simple configuration, and cost effectiveness. Every single in-home appliance generate particular data flow and may have specific communication requirement, but in general in-home wireless solution should be realized with multipath environment due to surface reflection and interference with other intelligent devices at home.
In near future it is expected that smart meter installed in the home connected to home appliances and to the utility control centers to control some appliances in order to reduce energy use and aggregate loads threw the AMI networks [38], this approach will allow customers to take advantages of lower tariff and optimized energy cost at off-peak time and reduce human intervention in energy cost optimization [39].

*Neighborhood Area Network (NAN)*
A NAN network is a distribution domain network, it can be considered as a mesh of smart meters. NAN connects the AMI applications access point to smart meters in customer domain and various gateways in the distribution domain. The main purpose of this network is data collection from smart meter for monitoring and control. It covers long distances up to 1–10 square miles and the data rate is around 10–1000 Kbps [12]. Both wired and wireless communication technologies could be appropriate for NAN networks, according to the authors in [17] for NAN networks WiMAX, LTE, 3G and 4G could be good candidates as a wireless communications technologies. While wired technologies such as PLC and Ethernet could be right solutions for NAN networks too. In NAN networks we can use either multi-hop or single hop approach liable on the technology deployed. We can give the example of WiMAX technology, the data from smart meters with WiMAX radio can be transmitted directly to the backhaul network and operation centers, a second approach is by transmitting the data from smart meters over multiple gateways before reaching the backhaul network.

*Field area Networks (FAN)*
Field Area Network is the communication network for distribution domain in the smart grid, the electrical power control centers and application use FAN networks to collect data, monitor and control different applications in distribution domain such as IED devices, PHEV charging stations, AMI applications in NAN networks and WSNs networks in feeders and transformers [2].

*Wide Area Network (WAN)*
WAN affords communications systems between smart grid and core utility system. It is composed by two types of networks backhaul and core network. The core network offers the connectivity between substations and utility systems, while backhaul network connect the NAN network to the core network, this network is extended over thousands of square miles and data rates reaches 10 to 100 Mbps [12]. A variety of technologies such as WiMAX, 4G, and PLC could be used in WAN networks. Also virtual technologies like IP/MPLS could be used for the core network.

# 5. Wired Communication technologies for smart grid

## 5.1. Power Line Communication (PLC)
The power line communication technology consists of introducing of the modulated carrier over the power line cable in order to establish two way communications [8], it is composed of two major categories Narrowband PLC and Broadband PLC, this technique permit utilities to utilize the power infrastructure to exchange data flows and monitoring control messages, and so far it is considered as a cost-effective smart grid communication means, and it is widely used in AMR applications deployment [9]. However PLC technology is deployed in several smart grid domains from bulk generation to distribution and end consumers. Therefore PLC can be considered as a practical solution for smart grid communication infrastructure [3]. In HAN environment PLC is not -until now- a suitable solution, because of the lack of interoperability and standards, the multi-protocol and the multi-vendor environment in HAN networks [8]. PLC technologies are preferred by utility operators because their reliability advantage compared with other communication techniques.

## 5.2. Fiber communication
Optical communication has been widely used to connect substations to operation and control centers in the backbone network [20] thanks to its multiple advantages such as robustness against radio and electromagnetic interferences making it a suitable choice for high voltage environs, and its capacity to transmit over large distances with very high bandwidth. We believe that Fiber optic communication will performance a crucial role in smart grid infrastructure, according to the authors in [21] the use of Optical Power Ground Wire (OPGW) technology in the distribution and transmission lines will be suitable in smart grid context since the combination of grounding and optical communications allow long distance transmissions with high data rates. Another application of fiber-optic technology would be to provide services to customer domain [22, 24] with the use of passive optical networks (PON) since they use only splitters to collect optical signals and do not require switching equipment. EPON for Ethernet PON is also interesting grid operators and seems to be suitable technology for smart grid access segment meanwhile its enable using interoperable IP-based Ethernet protocols over optical networks technology [6].

### 5.3. DSL

DSL for Digital subscriber lines, it is a suit of communication technologies permitting data transfer over telephone lines. Its main advantages consist of its simple utilization in the smart grid context since electric utilities can make immediate advantage of them without any extra cost for additional deployment. There are number of DSL alternatives like ADSL for Asymmetric DSL that supports up to 8 Mbps for downstream and 640 Kbps for upstream, the ADSL 2+ with up to 24 Mbps and 1 Mbps for downstream and upstream respectively. And VDSL (for Very high bit DSL) providing up to 52 Mbps for downstream and 16 Mbps for upstream but only for short distances [6].

## 6. Wireless Communication technologies for smart grid

### 6.1 IEEE 802.11 (WIFI)

IEEE 802.11 standard refers to the collection of wireless communication technology known as WIFI used for WLANs networks. This technology has proved its success due to its simple access structure based on CASMA/CA and its operation in unlicensed frequency bands (2.4 GHz and 5 GHz) [6]. IEEE 802.11 is a standards family; the latest release is the IEEE 802.11n which supports the highest data rates up to 150 Mbps while IEEE 802.11a/g supports maximum 54 Mbps. Other standard like 802.11e [23] appears to be important for SG applications because its QOS features, and the 802.11s standard allowing multi-hop and mesh networks over physical layer [24] and finally 802.11p standard for wireless networks for V2G systems[25].

### 6.2 IEEE 802.16 (WiMAX)

The IEEE 802.16 standard known as WiMAX supports long distance up to 10 Km broadband with up to 100Mbps of data rate [26]; WiMAX was designed to handle thousands of synchronized users over large distances. The 802.16j standard is the recent version of WiMAX supporting multicast and broadcast multi hop technique with seamless handover for mobile users [27] it empowers flexible distribution and higher coverage which make it suitable choice for NAN and AMI applications. The WiMAX under development version named 802.16 m will provide a greater mobility up to 350 km/h with 100 Mbps data rate [28], supporting handover with LTE and WIFI. The authors in [37] demonstrate with a simulation based on metering capacities and QOS that WiMAX technology is the most suitable than cellular and wired solutions in the distribution domain for the SG.

### 6.3. GSM, GPRS and EDGE

The cellular technology main advantages over wireless technologies is the larger coverage area, that why utilities have used them especially in AMR systems and SCADA [29, 30] but the high cost of this technology with problems such latency if a large number of users are served by the same base station has to be solved. Cellular technologies are endorsing great evolution in the few recent years with the development of 3G standards such high packet access standard (HSPA+) providing data rates up to 168 Mbps in the downlink and 22Mbps in the uplink[31].

### 6.4. Long Term Evolution (LTE)

The 4G standard or LTE for long term Evolution advanced is a wireless communication standard providing an enhancement of the LTE standard deployed today [30] introducing capabilities like bandwidth, easing handover between different networks and advanced networking proficiencies. LTE has multiple advantages that make it a good choice for NAN networks such as end to end quality of services, peak upload rates near 75 Mb/s, and download rates reaching 300 Mb/s [7]. The implementation of LTE technology in smart grid framework can be done with two ways, the first one is the most simple for immediate implementation, it is an efficient and a cost effective way it consist of carrying the data over the actual mobile network architecture of MNOs (Mobile network operators) with piggybacking technique from smart grid end devices in the HANs over the NAN network to the WAN until the utility. The second approach consist of utilization of a special network architecture for data transfer, and the implementation of this approach is similar to the MVNO (Mobile Virtual network operators) method it can be done by the rental of a portion of the MNO core network by the smart grid utility, or it can be done also by the implementation of the core network architecture by smart grid utility using the LTE technologies like the MNOs but totally disconnected from the MNO core network [7]. The cost effectiveness security and the simple implementation make LTE a good choice as a communication technique for NAN networks.

### 6.5 IEEE 802.15.4

The IEEE 802.15.4 is a standard for physical and MAC layers for low-rate wireless personal Area networks (LR-WPAN) it offers up to 250 kbps over 10m. Several network topologies are supported like star, tree or mesh multi-hop. IEEE 802.15.4 is the basis radios for many other standards for monitoring and control applications; the most important are ISA 100.11a standard, Zigbee standard and wirelessHART standard. These standards replace 802.15.4 MAC protocol with TDMA based scheme. Zigbee is widely adopted for WPANs for both commercial and industrial environments; further information about this standard will be given in section

### 6.6. ZigBee

ZigBee is a standard and a communication technology which specifies the physical layer and the medium access layer it is based on the IEEE 802.15.4 standard and commonly known as the low-rate wireless personal area networks (LR-WPANs). In ZigBee network we have two types of devices: FFD for full function devices and RFD for reduced function device, the FFD perform the network establishment, management and routing while RFD support FFD functionalities [10, 11].The network is composed of three types of nodes: the coordinator, the router and the end device [3].RFD is always the end device while FFD can make any device type. The role of the coordinator is to establish and manage the network, the router routes the traffic between coordinator and end device [11]. Coordinators and routers are battery powered devices, they are usually not allowed to go to sleep mode, and they are capable of communicating with all the rest of intelligent devices in the network. The end devices are permitted only to interconnect with the router or the coordinator and not with each other. End devices are awakening periodically to checks the parent node for their tasks and sends the data and then they go to sleep mode. The sleep mode in Zigbee network nodes make the network energy efficient and low power comparing with other communication technologies [11]. Zigbee most strong feature is its capability of creating application profiles empowering multi-vendor interoperability, in this profile we found a description of different parameters like data formats, application supported devices, message types etc. [6].

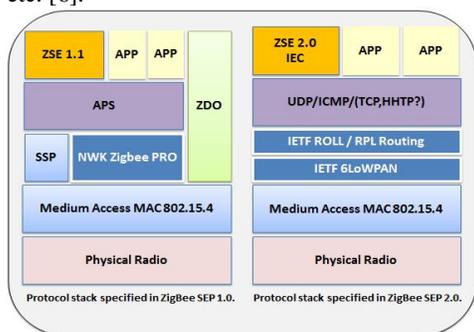

**Figure 4** Evolution of ZigBee Protocol Stack for SG new requirements

The ZigBee SEP 2.0 smart energy profile provides interfaces to manage, control and monitor energy use. As described by figure 4, the ZigBee IP protocol stack specified in SEP 2.0 compresses the packet structure specified in SEP1.0 within IPV6 packets. This will provide an independent interface between network and MAC layers tolerating smooth communication with IP based network.

### 6.7 Cognitive radio Networks

Cognitive radio technology is a stand-alone radio based on IEEE 802.22; it is a key technology for optimizing the underutilization of spectrum [32, 34] due to spectrum increasing demands caused by advancement of wireless technologies. CR networks enable to secondary users (SUs) the spectrum access when it is not used by the primary licensed user efficiently without causing any interference with PU. This spectrum sensing technique could be widely deployed in SG WAN, backhaul and distributions networks over large geographic area. The CR technique consists of opportunistic access to unused spectrum, we believe that this technique will have a great future for SG since it delivers a high performance, high-speed data transmission, scalability and fault tolerant broadband access. A Cognitive Radio Network Testbed is built in Tennesee technological university [35] in order to attain convergence between CR technique and Smart grid [36]. Cognitive radios make the smart grid "smarter" and provide to it more security, scalability, robustness, reliability and sustainability

### 6.8 DASH7

Dash 7 is a technology for wireless sensors networks based on ISO/IEC 18000-7 standard and promoted by Dash 7 Alliance. It is made for Active radio Frequency identification devices or RFIDs. Dash 7 operates at a 28 kbps rate up to 200 kbps and it has coverage of about 250 m extendable to 5 km [33]. It is a low power technology, with tiny sensors stacks and long live battery's up to several years which make it cost effective solution. The dash 7 uses very small amount of energy for wake up signal up to 30-60 mW and it is low latency with around 2.5-5 s [33].Its several advantages like interoperability, robustness and cost effectiveness made it widely deployed for military application and also commercial applications such building automation, smart energy, smart home, PHEVs, logistics control and monitoring. In the context of smart grid, Dash 7 seems to be respectable alternative to ZigBee, allowing several advantages like its wide range avoiding the multi-hop technique for HAN solution, permitting less number of nodes and less communication time.

## 7. Adopted End to end communication Architecture

The smart grid communication infrastructure will be composed of tiered level networks as described by figure 5 this architecture consists of three major domains which are the access segment, the distribution backhaul and the core network. Our objective in this section is to propose detailed end to end communication architecture using the reference model described in the literature [20, 6, 2 and 18].

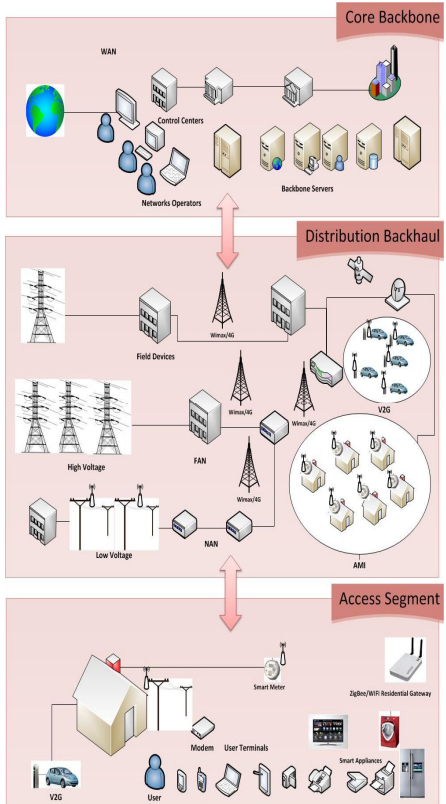

**Figure 5** End to End Smart Grid Communication Architecture

This architecture will provide a clear and a comprehensive view of communication scenarios between electrical devices in different smart grid domains and level with interconnection of small scale networks to form a large scale communication infrastructure providing high and secure connectivity between dispersed electrical devices in large national areas with different network topologies and hybrid but heterogeneous network technologies.

The smart grid networks in access tier level are responsible of data flows between customers and energy utilities and operation centers allowing an active and important role for end consumers. Whereas HAN are crucial elements in access tiers using wired , wireless or hybrid technologies and permitting monitoring and control of smart grid intelligent end devices at customers' premises in order to achieve energy conservation and wise usage of resources'. The key component is such networks are the HAN residential gateways designed with multiple radio interference tolerating the integration of different class of devices.

Communication networks in distribution backhaul are responsible for interconnection between smart grid core backbone and local area networks in access segment tier, they enable real time control and monitoring for distribution grid. The AMI networks, NAN and FAN networks will be considering both wireless and wired technologies for their deployment and options will include technologies such PLC WiMAX, LTE and WIFI among other possibilities.

The wide area network in core tier is require high-capacity communication technologies in order to deliver large amount of data from AMI systems and FAN networks to remote control centers. Using the public networks would be a good choice if challenges such latency, reliability and security are resolved. Another solution for electrical utilities is growing consisting of deployment of private WAN networks for that purpose using hybrid technologies between Fiber and wireless [22] in core backbone tier.

## 8. Smart grid Communication challenges

Smart grid communication most important challenges are interoperability, efficiency and performance. In order to permit the real deployment of smart grid systems different communication for utility companies, users and vendors should be adopted under a heterogeneous platform allowing dynamic and efficient coexistence of multiple equipment and techniques. Interdisciplinary is also a key component of this infrastructure, smart grid infrastructure will integrate different domains such as communication and networking technologies with security and power systems. The huge scale covered by SG systems involving millions of potential users -all over the globe- require a larger scalability and QOS between wired and wireless technologies. Security and privacy also will hinder electrical grid modernization because of the growing complexity of SG communication systems require novel techniques and measurements against unauthorized access and cyber vulnerabilities.

**Table 1**: Summary of communication technologies for Smart grid

| Family | Coverage | Applications | Advantages | Disadvantages |
|---|---|---|---|---|
| PLC | NB-PLC: 150 km BB-PLC: 1.5 km | NB-PLC: AMI, FAN, WAN BB-PLC: HAN/AMI | Communication infrastructure for SG is already established. Low costs. Separation from other communication networks. | Non-interoperable High signal attenuation Channel distortion Interference with electric appliances and electromagnetic sources High bit rates difficulties Complex Routing |

| Fiber | Between 10 Km and 60 Km depends on standard used | WAN AMI | Very Long-distance Ultra-high bandwidth Robustness against interference | High costs Difficult to upgrade Not suitable for metering applications |
|---|---|---|---|---|
| DSL | Between 300 m and 7 km depends on the variant used | AMI FAN | Commonly deployed for residential users Infrastructure is already established | High prices to use Telco networks Not suitable for long distances |
| WIFI | Between 300 m outdoors and up to 1 Km depending on versions | V2G HAN AMI | Low-cost network deployments and equipment's Flexibility Has several use cases | High interference High Power consumption Simple QoS support |
| WIMAX | Between 10 km and 100 km depends on performance | AMI FAN WAN | Suitable for high range of simultaneous Longer distances comparing with WiFi Connection-oriented Sophisticated QoS | Complex Network management High cost of terminal equipment Licensed spectrum use |
| 3G GSM,GPRS,EDGE | HSPA+: 0–5 km | V2G HAN AMI | Supports millions of devices Low power consumption Cellular SG specific service solutions. High flexibility, suitable for different use cases Licensed spectrum use reduces interference Open industry standards | Could have High prices for the use of Telco operators networks Licensed spectrum use Latency |
| 4G LTE/LTE-A | 0–5 km up to ~100 km (impacting performance) | V2G HAN AMI | Same as 3G with higher flexibility, enhanced technology, and handover. | Latency High cost |
| WPAN IEEE 802.15.4 | Between 10 and 75 m | V2G HAN AMI | Low power consumption Low cost Suitable for tiny devices with low resources | Low bandwidth Not suitable for large networks |
| ZigBee | Up to 100m | V2G HAN AMI | New ZigBee SEP 2.0 standards with full interoperability with IPv6 based networks | Low bandwidth Not suitable for large networks |
| Cognitive Radio | Up to 100 Km | AMI WAN | Very long distances High performance Scalable Fault tolerant broadband access Reliability | Interference with licensed users |
| Dash7 | Typical range about 250 m extendable to 5 km | V2G HAN AMI | Low cost Low power Lower range than ZigBee Efficiency | Not suitable for large networks |

## 9. CONCLUSION AND FUTURE DIRECTIONS

In this article, we have explored related networks and communication technologies that could be adopted for the smart grid communication infrastructure on smart grid distribution and customer domains. To make the right choice of communication technologies in smart grid communication networks, we have pinpointed their advantages and disadvantages for a variety of smart grid applications. Furthermore, we proposed an end to end communication infrastructure. We believe that the future power grid communication technologies choice is a critical issue for the purpose of enhancing and optimizing SG communications. Future work in the area includes looking at the routing protocols in communication networks for smart grid between in-home smart appliances, smart meter, AMI networks and the operator's control center. This will encompass reviewing a different class of protocol families for every network type in all smart grid segments. Such research will be complementary to the work presented here.

## AUTHORS

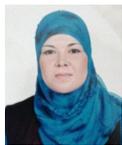

Saida Elyengui: is a PhD student in the department of communication systems at Tunisian National School of Engineering, University of Tunis El Manar Tunisia. She is a researcher in the area of smart grid communication and networking, SG networks security, AMI applications and M2M communications. She received her Computer Networks Engineer Diploma and a Master degree in new Technologies of Communication and Networking in 2007 and 2011 respectively.

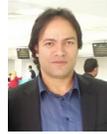

Riadh Bouhouchi: is an international IT and Communication systems consultant he received his Master degree in communication systems from ENIT, a T.E diploma in 2000 from ESPTT, and he holds an engineering Diploma in computer sciences since 2006, as he holds more than 8 international certificates in advanced programming and management as ITIL, he is also an active researcher in the area of sensors networks, SWE and M2M communications.